\documentstyle[12pt,epsfig]{ioplppt}

\begin{document}

\jl{4}

%%%%%%%%%%%%%%%%%%%%%%%%%%%%%%%%%%%%%%%%%%%%%%%%%%%%%%%%%%%%%%%%%%%%%%

\letter{Total spectrum of photon emission by an ultra-relativistic 
positron channelling in a periodically bent crystal.
\footnote{published in
J. Phys. G: Nucl. Part. Phys. {\bf 26} (2000) L87--L95, Copyright 2000
IOP Publishing Ltd., http://www.iop.org
}
}

\author{
Wolfram Krause\dag\ftnote{4}{E-mail: krause@th.physik.uni-frankfurt.de},
Andrei V. Korol\dag\ddag\ftnote{5}{E-mail:
korol@th.physik.uni-frankfurt.de,\ korol@rpro.ioffe.rssi.ru},
Andrey V. Solov'yov\dag\S\ftnote{6}{E-mail: 
solovyov@th.physik.uni-frankfurt.de,\  solovyov@rpro.ioffe.rssi.ru},
and Walter Greiner\dag\ftnote{7}{E-mail: greiner@th.physik.uni-frankfurt.de}
}

\address{\dag Institut f\"ur Theoretische Physik der Johann Wolfgang
Goethe-Universit\"at, 60054 Frankfurt am Main, Germany}

\address{\ddag Department of Physics,
St.Petersburg State Maritime Technical University,
Leninskii prospect 101, St. Petersburg 198262, Russia}

\address{\S A.F.Ioffe Physical-Technical Institute of the Academy
of Sciences of Russia, Polytechnicheskaya 26, St. Petersburg 194021,
 Russia}


% PACS numbers:
%
% Electromagnetic radiation
%  -from moving charges, 41.60
%
% Undulator radiation, 41.60

%%%%%%%%%%%%%%%%%%%%%%%%%%%%%%%%%%%%%%%%%%%%%%%%%%%%%%%%%%%%%%%%%%%%%%

%%%%%%%%%%%%%%%%%%%%%%%%%%%%%%%%%%%%%%%%%%%%%%%%%%%%%%%%%%%%%%%%%%%%%%

\begin{abstract}
We present the results of numerical calculations of the channelling
and undulator radiation generated by an ultra-relativistic positron
channelling along a crystal plane, which is periodically bent.  The
bending might be due either to the propagation of a transverse
acoustic wave through the crystal, or due to the static strain as it
occurs in superlattices.  The periodically bent crystal serves as an
undulator.  We investigate the dependence of the intensities of both
the ordinary channelling and the undulator radiations on the
parameters of the periodically bent channel with simultaneous account
for the dechannelling effect of the positrons.  We demonstrate that
there is a range of parameters in which the undulator radiation
dominates over the channelling one and the characteristic frequencies
of both types of radiation are well separated.  This result is
important, because the undulator radiation can be used to create a
tunable source of X-ray and $\gamma$-radiation.
\end{abstract}

%%%%%%%%%%%%%%%%%%%%%%%%%%%%%%%%%%%%%%%%%%%%%%%%%%%%%%%%%%%%%%%%%%

In this Letter we present for the first time the results of
calculations of the total spectrum of emitted photons accompanying the
channelling process of ultra-relativistic charged particles through a
crystal which is periodically bent.  The present consideration is a
further step in investigating new phenomenon which was described
recently in \cite{JPG,laser} and was called Acoustically Induced
Radiation (AIR).  It was noted that the periodic pattern of crystal
bendings (which can be achieved either through propagation of a
transverse acoustic wave, AW, or by using static periodically strained
crystalline structures \cite{laser,Uggerhoj2000}) gives rise to a new
mechanism of electromagnetic emission of the undulator type, in
addition to a well-known ordinary channelling radiation
\cite{Kumakhov}.

\begin{figure}
\hspace{2.7cm}\epsfig{file=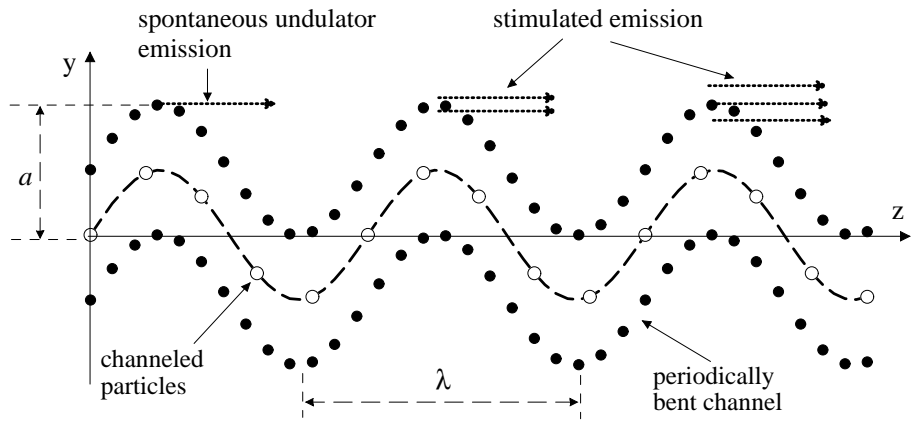, width=12cm}
\caption{Schematic representation of spontaneous and stimulated 
AIR  in a channel bent by the transverse AW. 
The $y$- and $z$-scales are incompatible.}
\label{Fig.1}
\end{figure}

Without any loss of generality further in this Letter we consider the
case of a dynamic periodic bending of a crystal by means of transverse
AW, as it is illustrated in \fref{Fig.1}.  Under the action of the
transverse AW propagating along the $z$-direction, which defines the
center line of initially straight channel (not plotted in the figure)
the channel becomes periodically bent.  Provided certain conditions
are fulfilled \cite{laser} the beam of particles, which enters the
crystal at a small incident angle with respect to the curved
crystallographic plane, will penetrate through the crystal following
the bendings of its channel.  This results in transverse oscillations
of the beam particles ({\it additional} to the oscillations inside the
channel due to the action of the interplanar force).  These
oscillations become an effective source of spontaneous radiation of
undulator type due to the constructive interference of the photons
emitted from similar parts of the trajectory.  It was demonstrated
\cite{laser} that the system ``ultra-relativistic charged particle +
periodically bent crystal'' serves as a new type of undulator, and,
consequently, as a new source of undulator radiation of high
intensity, monochromaticity and of a particular pattern of the
angular-frequency distribution.

%%% Fig.1

As it was pointed out in \cite{laser} this scheme leads, in addition
to the spontaneous radiation, to a possibility to generate stimulated
emission, similar to the one known for a free electron laser
\cite{Madey71} in which the periodicity of a trajectory of an
ultra-relativistic projectile is achieved by applying spatially
periodic magnetic field.  In connection with the stimulated AIR it was
noted in \cite{laser} that to achieve noticeable degree of
amplification one has to operate with a positron bunch of a high
volume density, which, nevertheless, turned out to be achievable in
modern accelerators as discussed in \cite{Zhang1999}.

The main subject of our previous studies \cite{JPG,laser} was the
characteristics of the undulator AIR itself.  Due to this reason we
primarily investigate the case of a high-amplitude AW, $a\gg d$ with
$d$ standing for the interplanar spacing, propagating through the
crystal.  Less attention, except for some qualitative estimates, has
been paid to the detailed investigation of the mutual influence of two
types of radiation, the ordinary channelling radiation and the AIR, in
forming the total spectrum of the radiation emitted by a channeled
positron.  This topic is addressed in the present Letter where we
report our first quantitative results on numerical calculation of the
spectrum of the emitted photons with both mechanisms taken into
account simultaneously.  We demonstrate that there are ranges of (i)
the parameters of AW, which are the amplitude, $a$, the wavelength
$\lambda$, and the sound velocity $V$, (ii) the energies of projectile
positron, $\varepsilon$, (iii) the crystal parameters, which include
the length of a crystal, the constituent atoms and the types of
crystallographic planes, inside which
\begin{itemize}

\item the characteristic frequencies of the AIR and the ordinary channelling
radiation are well separated,

\item the intensity of AIR is essentially higher than that of the ordinary
channelling radiation,

\item the radiative spectrum is stable towards the total losses of the
particle (in the case of a positron it is primarily the radiative
ones)
\end{itemize}

These items, except the last one which was considered in
\cite{losses}, are discussed below in the Letter.

To conclude the introductory part we mention that in the past the
problem of evaluating the total spectrum of radiation formed in a bent
crystal was considered in several publications
\cite{Taratin,Arutyunov} in the case of a projectile channelling in a
crystal bent with a constant curvature radius.  In our Letter for the
first we investigate the problem for the periodically bent channel.
In full the results of our research will be published elsewhere
\cite{new_full}. Below we present the essential points.

An adequate approach to the problem of the radiation emission by an
ultra-relativistic particle moving in an external field was developed
by Baier and Katkov in the late 1960s \cite{Baier67} and was called
``operator quasi-classical method'' by the authors.  The details of
that formalism can be found in \cite{Baier,Land4}.  The advantage of
this method is that it allows to use the classical trajectory for the
particle in an external field and, simultaneously, it takes into
account the effect of the radiative recoil.

For particles with spin $s=1/2$ the energy radiated into a given
direction ${\bf n}$ summed over the polarizations of the photon and
the projectile is given by (the CGS system is adopted throughout the
paper)
\begin{equation}
\d E \equiv
{\d E \over \hbar\d \omega\, \d \Omega_{\bf n} }=
\frac{\alpha\, \omega^2}{4 \pi^2}
\int_0^\tau \d t_1 \int_0^\tau \d t_2 \
\e^{\i \omega^{\prime} \varphi(t_1,t_2)}
\ f(t_1,t_2),
\label{wkb_1}
\end{equation}
where $\alpha\approx 1/137$ is the fine structure constant,
$\varphi(t_1,t_2)=t_1-t_2- {\bf n}\cdot({\bf r}(t_1)-{\bf r}(t_2))/c$,
and $f(t_1,t_2)=\left\{\left[ 1+(1+u)^2\right] \left({\bf v}(t_1){\bf
v}(t_2)/c^2-1\right)+u^2\gamma^{-2}\right\}/2$, where $c$ is the
velocity of light, $\gamma=\varepsilon/m$ is the relativistic factor
and $u=\hbar \omega/(\varepsilon - \hbar \omega)$.

The main advantage of the the quasi-classical approach is that to
calculate the angular-spectral distribution of the radiation one only
needs to know the time dependencies of classical radius-vector ${\bf
r}(t)$ and the velocity ${\bf v}(t)$ of projectile.  In connection
with a positron channelling through a periodically bent crystal the
above relations impose, if applied directly, some restrictions on the
projectile energy, on the parameters of the crystal (channel) and the
AW.
  
The channelling process in a bent crystal takes place if the
centrifugal force in the channel is less than the maximal force due to
the interplanar field \cite{Tsyganov}.  For a periodically bent
crystal the maximal centrifugal force is equal to $m \gamma v^2/R_{\rm
min}$, with $v\approx c$ and $R_{\rm min}$ being a minimum curvature
radius of the bent channel.  Hence, the following condition must be
fulfilled \cite{JPG,laser}
\begin{equation}
m \gamma c^2/R_{\rm min} < U_{\rm max}^{\prime}.
\label{1}
\end{equation}
where $U_{\rm max}^{\prime}$ stands for the maximum gradient of the
interplanar field.  For an acoustically bent channel $R_{\rm min}
=(\lambda/2\pi)^2/a$, therefore, the inequality (\ref{1}) bounds
together the characteristics of the crystal, $U_{\rm max}^{\prime}$,
the AW amplitude and the wavelength, and the relativistic factor
$\gamma$.  A comprehensive study of the allowed ranges of all
parameters involved in (\ref{1}) was carried out in \cite{JPG,laser}.

Second requirement which has to be fulfilled to make eq. (\ref{wkb_1})
directly applicable to calculating the spectrum accompanying the
channelling process concerns the upper limit of integration $\tau$,
which is related to the crystal length $L$ through $L=c\tau$, and
which, for given value of the AW wavelength, defines the number of the
undulator periods as $N=L/\lambda$.  The AIR acquires specific
features of the undulator-type radiation (such as the monochromaticity
and particular pattern of the angular-frequency distribution, see
e.g. \cite{Baier}) provided $N\gg 1$.  The length of a crystal is
subject to a physical condition that the positron bunch stays inside
the channel when penetrating through the crystal on the scale of $L$.
In reality, the parasitic effect, the dechannelling leads to a
decrease in the volume density of the channeled particles $n(z)$ with
penetration distance $z$, and roughly satisfies the exponential decay
law for both straight and bent channels (see \cite{Gemmell,Biryukov}),
$n(z) = n_0\, \exp\left(- z/ L_d\right)$, where $n_0$ is the volume
density t the entrance, and $L_d$ is the dechannelling length, which
for given crystal and channel depends on a positron energy and on the
curvature radius.  For a periodically bent crystal $L_d(\gamma, R)$
can be estimated as
\cite{laser,Biryukov,Komarov}
\begin{equation}
L_d(\gamma, R)
= 
\left(1 - R_c/R_{\rm min}\right)^2 \,
{256 \over 9\pi^2}\,{ a_{\rm TF} \over r_{\rm cl} }\,
{ d \over L_c }\, \gamma
\label{stim9}
\end{equation}
where $r_{\rm cl}$ is the classical radius of an electron, $a_{\rm
TF}=0.8853 Z^{-1/3} a_0$ ($a_0$ is the Bohr radius) and $I = 16
Z^{0.9}$ eV are, respectively, the Thomas-Fermi radius and ionization
potential of the crystal atoms, $Z$ is the atomic number, $d$ is the
interplanar distance in the lattice.  The quantity $R_c= \varepsilon/
U_{\rm max}^{\prime}$ is the critical (minimal) radius consistent with
the channelling condition in a bent crystal (\ref{1}).  The quantity
$L_c =\ln\left(\sqrt{2\gamma}\, mc^2/I\right) - 23/24$ is the Coulomb
logarithm characterizing the ionization losses of an
ultra-relativistic positron in amorphous media with account for the
density effect (see e.g. \cite{Komarov,Sternheimer}).

The AIR will have a pronounced pattern of undulator-type radiation
provided the number of the periods is large on the scale of $L_d$.
Combined with (\ref{stim9}) this condition leads to another
restriction on the parameters involved
\begin{equation}
N = \left(1 - R_c/R_{\rm min}\right)^2 \,
{256 \over 9\pi^2}\,{ a_{\rm TF} \over r_{\rm cl} }\,
{ d \over L_c }\, {\gamma \over \lambda} \gg 1
\label{2}
\end{equation}
where $\lambda$ can be related to the AW frequency $\nu$ through
$\lambda=V/\nu$ with $V$ standing for the sound velocity.

\begin{figure}% 
\hspace{2.7cm}\epsfig{file=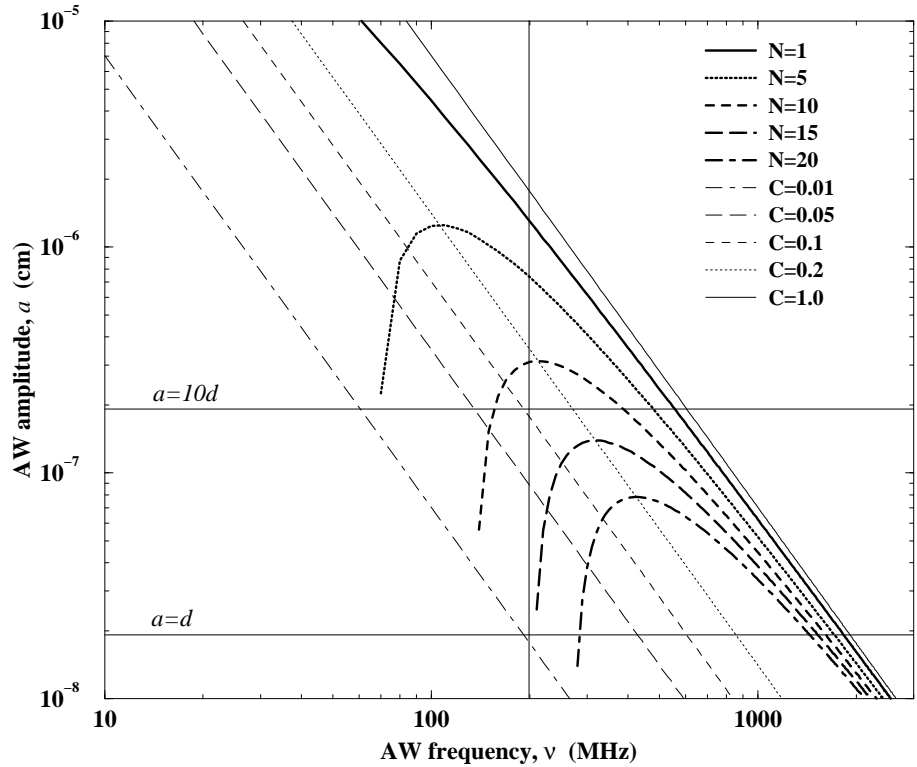, height=12cm, angle=270}
\caption{
The ranges of the allowed parameters of the AW, $a$ versus $\nu$, 
consistent with the conditions (\ref{1}) and (\ref{2})
for $\varepsilon=0.5$ GeV ($\gamma \approx 10^3$) positron 
channeling in $Si$ along (110) crystallographic planes.
See explanations in the text for the meaning of various lines.
}
\label{Fig.2}
\end{figure}

\Fref{Fig.2} illustrates the restrictions which are imposed
on the values of $a$ and $\nu$ by inequalities (\ref{1}) and (\ref{2})
in the case of $\varepsilon=0.5$ GeV planar channelling in $Si$ along
(110) crystallographic planes.  The diagonal straight lines correspond
to various values (as indicated) of the parameter
$C=\varepsilon/(R_{\rm min}U_{\rm max}^{\prime}) \leq 1$ consistent
with the channelling condition (\ref{1}).  The curved lines correspond
to various values (as indicated) of the number of undulator periods
$N$ related to the dechannelling length $L_d$ through eq. (\ref{2}).
The horizontal lines mark the values of the AW amplitude equal to $d$
(with $d=1.92\, \AA$\, being the (110) interplanar distance in $Si$)
and to $10\,d$.  The vertical line marks the value $\nu = 200$ MHz of
the AW frequency for which the spectra (\ref{wkb_1}) were calculated.
We used the value $V=4.67\times 10^5$ cm/s for the velocity of sound
in $Si$ (this value was obtained by using the data from
\cite{Mason}). Thus, the AW wave-length used in our calculations
equals to $\lambda=2.33\times 10^{-3}$ cm. 

%%% Fig.2

\begin{figure}% 
\hspace{2.7cm}\epsfig{file=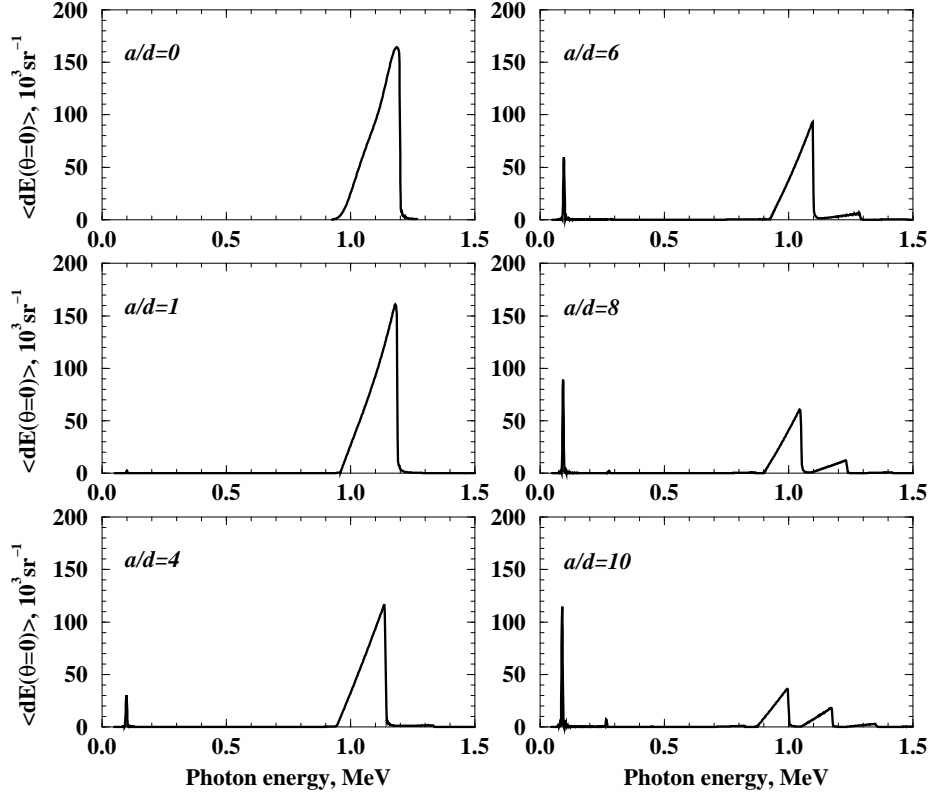, width=12cm}
\caption{
Spectral distribution (\ref{3})
of the total radiation emitted in the forward
direction (${\bf n}\, \| z$) 
for $\varepsilon=0.5$ GeV ($\gamma \approx 10^3$) positron 
channeling in $Si$ along (110) crystallographic planes 
calculated at different $a/d$ ratios as indicated.
The AW frequency is fixed at $\nu = 200$ MHz, 
the crystal length is $L=3.5\times 10^{-2}$ cm.}
\label{Fig.3}
\end{figure}

The calculated spectra of the radiation emitted in the forward
direction (with respect to the $z$-axis, see \fref{Fig.1}) for photon
energies from 45 keV to 1.5 MeV are presented in figures \ref{Fig.3}.
The details of the analytical evaluation of the right-hand side of
(\ref{wkb_1}) and its numerical implementation as well as more
extended results of the calculation will be published soon
\cite{new_full}.  Here we briefly sketch the numeric procedure and
present a discussion of the exhibited results.  The AW frequency, the
number of undulator periods and crystal length were fixed at $\nu =
200$ MHz, $N=15$ and $L=N\, \lambda = 3.5\times 10^{-2}$ cm.  The
ratio $a/d$ of the AW amplitude to the interplanar spacing was varied
within the interval $a/d = 0\dots 10$.  The case $a/d=0$ corresponds
to the straight channel.

%%% Fig.3

To evaluate the spectral distribution (\ref{wkb_1}) the following
procedure was adopted.

Firstly, the spectrum was calculated for individual trajectories.
These were obtained by solving the relativistic equations of motion
with both the interplanar and the centrifugal potentials taken into
account.  We used the continuum approximation \cite{Lindhard} to
describe the projectile positron -- lattice atoms interaction.  Within
these scheme we considered two frequently used \cite{Gemmell} analytic
forms for the interplanar potential, the harmonic and the Moli\`ere
potentials calculated at the temperature $T=150$ K to account for the
thermal vibrations of the lattice atoms.  For each $a/d$ value by
changing the initial values of the entrance coordinate $y^{(0)}$ and
the initial velocities along the $y$-axis (see \fref{Fig.1}) we left
only those trajectories which corresponded to the case of stable
channelling through the whole crystal length $L$.  We call a
trajectory as a ``stable'' one if moving along it the particle does
not approach crystalline planes at a distance less than the
Thomas-Fermi radius $a_{\rm TF}$ ($a_{\rm TF}=0.194$ \AA\, for a $Si$
atom).  This allowed us to totally disregard, at least on the scale
$L\sim L_d$, the random scattering of a projectile by lattice
electrons (see e.g. \cite{Gemmell,Biryukov}).  Thus, for each $a/d$
value, we defined the ranges of the initial coordinates $y^{(0)}\in
[-d/2+a_{\rm TF}, d/2-a_{\rm TF}]$ and the velocities $v_y^{(0)}$,
and, correspondingly, the initial phase volume $\Phi^{(0)}(a/d)=\oint
p_y^{(0)} \d y^{(0)}$ (where $p_y^{(0)}$ stands for initial transverse
momentum) for which the corresponding classical trajectories are
stable when channelling through the whole crystal length $L$.  Then,
discretizing the found initial phase volume $\Phi^{(0)}(a/d)$ by
choosing $N_{y^{(0)}}\times N_{v_y^{(0)}}$ points $(y^{(0)},
p_y^{(0)})\in \Phi^{(0)}(a/d)$, the individual spectra $\d E(y^{(0)},
p_y^{(0)})$ were calculated for each pair of the initial coordinate
and velocity.  Finally, for each $a/d$ value we evaluated the averaged
spectrum defined as follows:
\begin{equation}
\langle \d E \rangle
=
{1 \over \Phi^{(0)}(a/d=0)}\,
\oint_{\Phi^{(0)}(a/d)} \left[\d E(y^{(0)}, p_y^{(0)})\right]\,
p_y^{(0)} \d y^{(0)}
\label{3}
\end{equation}
Here, the integration is carried out over the phase volume
$\Phi^{(0)}(a/d)$, and the integral is scaled by the phase volume
$\Phi^{(0)}(a/d=0)$ of stable trajectories in straight channel.  The
ratio $\Phi^{(0)}(a/d)/\Phi^{(0)}(a/d=0)$ describes the number of
particles channelled through the bent crystal relative to the number
of particles channelled through the straight one.  Hence, it is
convenient to use the quantity $\langle \d E \rangle$ to compare the
spectra produced by effectively different number of projectiles as it
happens for different $a/d$ values.

Figures \ref{Fig.3} correspond to the spectra (\ref{3}) obtained as
outlined above. The results presented were calculated by using the
Moli\`ere approximation for interplanar potential.

The first graph in \fref{Fig.3} corresponds to the case of zero
amplitude AW (the ratio $a/d=0$) and, hence, presents the spectral
dependence of the ordinary channelling radiation only.  The asymmetric
shape of the calculated ordinary channelling radiation peak, which is
due to the strong anharmonic character of the Moli\`ere potential,
bears close resemblance with the experimentally measured spectra
\cite{Uggerhoj1993}. The spectrum starts at $\hbar\omega\approx 960$
keV, reaches its maximum value at $1190$ keV, and steeply cuts off at
$1200$ keV.  This peak corresponds to the radiation into the first
harmonic of the ordinary channelling radiation (see
e.g. \cite{Kumakhov1989}), and there is almost no radiation into
higher harmonics.  The latter fact is consistent with general theory
of dipole radiation by ultra-relativistic particles undergoing
quasi-periodic motion (see e.g. \cite{Baier}).  Dipole approximation
is valid provided the corresponding undulator parameter $p_c = 2\pi
\gamma(a_c/\lambda_c)$ is much less than 1.  In this relation $a_c$
and $\lambda_c$ stand for the characteristic scales of,
correspondingly, the amplitude and the wave-length of the
quasi-periodic trajectory.  For the channelling motion one can
estimate $a_c$ as $d/2$, and $\lambda_c=c \tau_c$, where $\tau_c\sim
2\pi\sqrt{m\gamma/U^{\prime\prime}}$ standing for the characteristic
period of the channelling oscillations (using the harmonic
approximation for the interplanar potential one gets $U^{\prime\prime}
\sim 8\, U_0/d^2$ where $U_0$ is the depth of the interplanar
potential well).  In the case of 0.5 GeV positron channeled along
(110) planes in $Si$ one has $\gamma\approx 10^3$, $U_0 = 23$ eV,
$d=1.92$ \AA\ \cite{Baier}.  Hence, $p_c\approx 0.2 \ll 1$ and all the
channelling radiation is concentrated within some interval in the
vicinity of the energy of the first harmonic.  The latter one can
estimate as (see e.g. \cite{Baier})
$\omega_c^{(1)}\sim4\pi\gamma^2/\tau_c\sim 4\gamma^2
c/d\sqrt{U_0/\varepsilon}$ arriving at the value
$\hbar\omega_c^{(1)}\approx 1190$ keV which is exactly the calculated
maximum value $1190$ keV.  More accurate estimates (the details are
presented in \cite{new_full}), based on the account for the dependence
of the channeling oscillation period $\tau_c$ on the amplitude $a_c$
of the oscillations, also reproduce the calculated position and the
width of the peak of the channeling radiation.

Increasing the $a/d$ ratio leads to the modifications in the spectrum
of radiation.  The changes which occur manifest themselves via three
main features, (i) the lowering of the ordinary channelling radiation
peak, (ii) the gradual increase of the intensity of undulator
radiation due to the crystal bending, (iii) the appearing of
additional structure (the sub-peaks) in the vicinity of the first
harmonic of the ordinary channelling radiation.  Let us discuss all
these features.

The decrease in the intensity of the ordinary channelling radiation
with the increase of the $a/d$ ratio is related to the simple fact
that the growth of the AW amplitude leads to lowering of the allowed
maximum value of the channelling oscillations amplitude $a_c$ (this
are measured with respect to the centerline of the bent channel)
\cite{losses}.  Indeed, inside the bent channel the motion of the
particle is subject to the action of the effective potential
\begin{equation}
U_{eff}(\rho) = U(\rho) - {\varepsilon \over R(z)}\, \rho
\label{2.25}
\end{equation}
where $\rho\in [-d/2,d/2]$ is the (local) distance from the centerline
and $R(z)$ is the (local) curvature radius of the channel.  For a
channel bent by a transverse harmonic AW $R =
\left[(\lambda/2\pi)^2/a\right]\,\sin(2\pi z/\lambda)$.  The particle
could be trapped into the channelling mode provided its total energy
associated with the transverse motion is less the minimal value,
$U_{eff}(\rho_0)$, of the two maxima points of the asymmetric
potential well described by (\ref{2.25}) \cite{losses,Biryukov}.  The
potential $U_{eff}(\rho)$ reaches the magnitude of $U_{eff}(\rho_0)$
at some point $\rho_0$ which satisfies the condition $|\rho_0|< d/2$
and the absolute value of $\rho_0$ decreases with the growth of $a$.
Hence, the larger the channel is bent the lower the allowed values of
the channelling oscillations amplitude are, and, consequently, the
less intensive is the channelling radiation, which is, essentially,
proportional to the squared amplitude \cite{Baier}.  Let us note, that
since we have restricted the range of the $a_c$ values by imposing the
condition $a_c < d/2-a_{TF}$ (see the discussion above \eref{3}), then
the decrease of the intensity of channelling radiation occurs starting
with some non-zero value of the AW amplitude for which $|\rho_0|$ also
becomes less than $d/2-a_{TF}$.

These arguments explain the fact that the intensities of the ordinary
channelling radiation for $a/d=0$ and $a/d=1$ are much alike, while
for $a/d>1$ it starts loosing the magnitude.

The undulator radiation (the AIR) related to the motion of the
particle along the centerline of periodically bent channel is absent
in the case of straight channel (the graph $a/d=0$), and is almost
invisible for comparatively small amplitudes of the AW (see the graph
for $a/d=1$).  With $a$ increasing the peaks corresponding to AIR are
becoming more prominent, and for large $a$ values ($a/d \sim 10$) two
additional features appear: the intensity of the first harmonic of the
AIR becomes larger than the intensity of the ordinary channelling
radiation, and there appears radiation into the third harmonic of the
AIR.

The positions and the widths of the AIR peaks can be quite accurately
estimated as follows (see \cite{JPG,laser}): $\omega^{(n)} = 8\pi
\gamma^2 c \lambda^{-1}\, n /(2+p^2)$, where the integer $n=1,2\dots$
enumerates the harmonics, and $p=2\pi\gamma(a/\lambda)$ stands for the
parameter of the undulator related to the periodicity of the channel
bending.  The width of each peak $\Delta \omega=
(1/N)(\omega^{(n)}/n)$ is independent on $n$.  Substituting the values
of $\gamma$, $\lambda$ and $d$ indicated above one expresses the
undulator parameter via the ratio $a/d$: $p\approx 0.05 (a/d)$.  Even
for the largest considered value $a/d=10$ the parameter $p$ is less
than 1, thus making the radiation into higher harmonics of the AIR
almost negligible compared with the intensity radiated into the
fundamental harmonic $n=1$.  The latter is located at
$\hbar\omega^{(1)} \approx 90$ keV having the width $\hbar\Delta
\omega \approx 6$ keV which is almost 40 times less than the width of
the peak of the channeling radiation.  These values depend neither on
the ratio $a/d=10$ nor on the type of the interplanar potential.

As mentioned, all graphs in \fref{Fig.3} refer to the forward
emission.  Therefore, in accordance with general theory of the
undulator radiation (see e.g. \cite{Baier}), the second peak of the
AIR, which is mostly pronounced in the case $a/d=10$, corresponds to
the third harmonic of AIR, and is located at $\hbar\omega^{(2)}
\approx 270$ keV.  The intensities radiated into the fundamental and
the third harmonics are equal to $1.10\times 10^{5}$ sr$^{-1}$ and
$7.12\times 10^{3}\ \mathrm{sr}^{-1}$, respectively.  Their ratio is
approximately equal to $p^{-4}$ which is also in accordance with
general theory.

The intensity of the AIR gradually increases with the AW amplitude.
More precisely, $\d E_{\rm AIR} \propto p^2\propto (a/d)^2$ (in the
case $p<1$), and this tendency one can observe comparing the AIR peaks
in the graphs corresponding to $a/d=4,6,8,10$.

\begin{figure}% 
\hspace{2.7cm}\epsfig{file=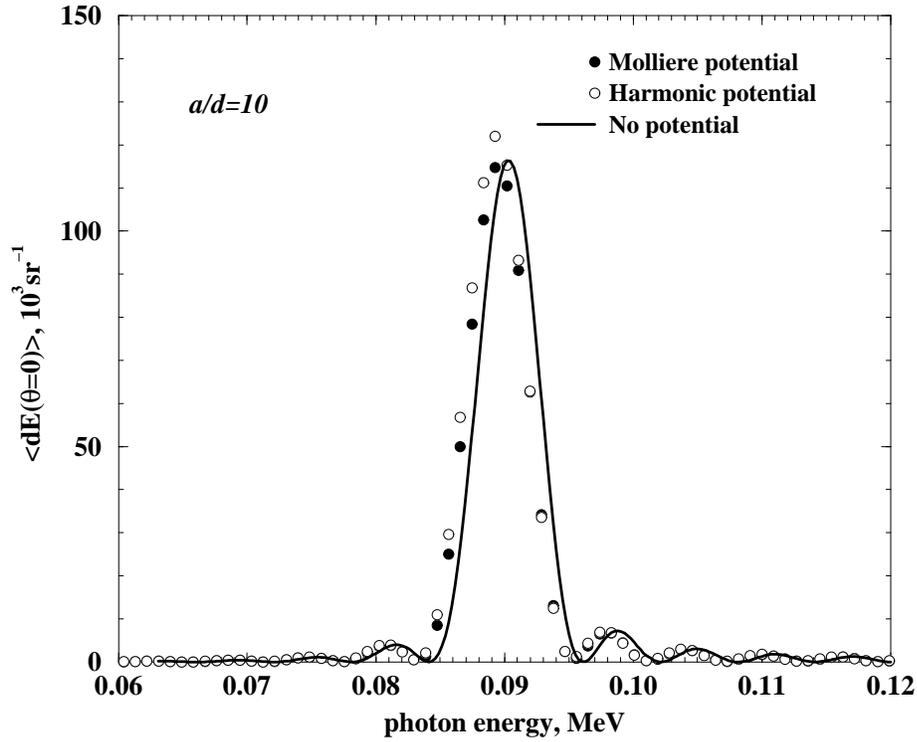, width=12cm}
\caption{
Comparison of different approximations used to calculate the total
ardiative spectrum in vicinity of the first harmonic of the AIR.
The ratio $a/d=10$, other parameters as in \fref{Fig.3}.
See also the commentaries in the text.
}
\label{Fig.4}
\end{figure}

It is important to note that the positions of sharp AIR peaks, their
narrow widths, and the radiated intensity are, practically,
insensitive to the choice of the approximation used to describe the
interplanar potential.  In addition, provided the condition (\ref{1})
is fulfilled, the AIR peaks are well separated (in the photon energy
scale) from the peaks of the channeling radiation.  Therefore, if
being interested in the spectral distribution of the AIR only, one may
disregard the channeling oscillations and to assume that the
projectile moves along the centerline of the bent channel
\cite{JPG,laser}. The above statements are illustrated by \fref{Fig.4}
where we compare the results of calculations of the total spectrum
(\ref{3}) in vicinity of the first harmonic of AIR in the case
$a/d=10$.  All parameters are the same as in \fref{Fig.3}.  The filled
and open circles represent the results of evaluation of the right-hand
sides of (\ref{1}) and (\ref{3}) accompanied by numerical solution of
the equations of motion for the projectile within the Moli\`ere
(filled circles) and the harmonic (open circles) approximations for
the interplanar potential.  The solid line corresponds to the AIR
radiation only (in this case the numerical procedures are simplified
considerably, leading to the reduction, by orders of magnitude, of the
CPU time).  It is clearly seen that the more sophisticated treatment
almost does not change the profile of the peak obtained by means of
simple formulae describing purely AIR radiation \cite{JPG,laser}.
Moreover, the minor changes in the position and the height of the peak
can be easily accounted for within the framework of the cited
formalism by introducing the effective undulator parameter
\cite{losses,new_full} and (in the case of the harmonic approximation)
the effective undulator amplitude \cite{new_full}.

%%% Fig.4

Thus, in vicinity of the AIR peaks there is no coupling of the two
mechanisms of the radiation.  On the contrary, the AIR strongly
affects the spectrum in the photon energy range corresponding to the
channeling radiation.  The discussion of this effect lies beyond the
scope of this Letter and is presented in \cite{new_full}.  Here we
only mention that, as it is clearly seen from the graphs $a/d=6,8,10$
in \fref{Fig.3}, for large values of $a/d$ the peaks of the ordinary
channeling radiation acquire additional structure: there appear
sub-peaks separated (in the case of the forward emission) by the
interval $\delta \omega = 2\omega^{(1)}$.

The presented results of the numerical calculations of the total
spectrum of radiation formed in an acoustically bent crystal clearly
demonstrate the validity of the statement made in
\cite{JPG,laser,losses} that the AIR and the ordinary channeling
radiation occur in essentially different ranges of the emitted photons
energies, allowing, thus, to investigate the AIR properties separately
from the ordinary channeling radiation.

The positions of the peaks of the AIR as well as their intensity can
be easily varied by changing the energy of projectile, the crystal
type, the type of crystallographic plane, and the parameters of
periodic pattern of crystal bendings.  This statement by no means is
restricted to the case of the acoustically bent crystal, but has more
general nature.  The treatment and the main results presented above
can be applied, in particular, to consider the undulator radiation in
statically bent crystals
\cite{laser,Uggerhoj2000}. 
Whatever way is chosen to prepare periodically bent crystalline
lattice it is worth investigating, both theoretically and experimentally, not 
only the spontaneous emission of the AIR-type, but the stimulated
emission as well \cite{laser}.

%%%%%%%%  Acknowledgments
\ack

The authors express their gratitude to Professor E. Uggerh{\o}j for 
sending a copy of the paper \cite{Uggerhoj2000} prior to its publication. 
We want to thank the administrative staff of the
Physics CIP Computer Cluster at University of Munich
for the possibility of using some of their computers for
numerical calculations.
The research was supported by DFG, GSI, and BMBF.
AVK and AVS acknowledge the support from the 
Alexander von Humboldt Foundation.

%%%%%%%%%%%%%%%%%%%%%%%%%%%%%%%%%%%%%%%%%%%%%%%%%%%%%%%%

\end{document}